# National research assessment exercises: a comparison of peer review and bibliometrics rankings[1]

*Giovanni Abramo*[a,b,*]*, Ciriaco Andrea D'Angelo*[b] *and Flavia Di Costa*[b]

[a] National Research Council of Italy, IASI-CNR

[b] Laboratory for Studies of Research and Technology Transfer
School of Engineering, Dept of Management
University of Rome "Tor Vergata"

**Abstract**

Development of bibliometric techniques has reached such a level as to suggest their integration or total substitution for classic peer review in the national research assessment exercises, as far as the hard sciences are concerned. In this work we compare rankings lists of universities captured by the first Italian evaluation exercise, through peer review, with the results of bibliometric simulations. The comparison shows the great differences between peer review and bibliometric rankings for excellence and productivity.

**\* Corresponding author:** Dipartimento di Ingegneria dell'Impresa, Università degli Studi di Roma "Tor Vergata", Via del Politecnico 1, 00133 Rome - ITALY, tel. +39 06 72597362, giovanni.abramo@uniroma2.it

## 1. Introduction

In recent years there has been unanimous agreement that governments should assign resources for scientific development according to rigorous evaluation criteria. This responds to the needs of the knowledge economy, which demands development of efficient scientific infrastructure capable of supporting the competitiveness of the national production system. The rising costs of research and tight restrictions on budgets add to the tendency for evaluation. Governments thus resort to such exercises, for the following purposes: i) to stimulate greater efficiency in research activity; ii) to allocate resources in function of merit; iii) to reduce information asymmetry between supply and demand for new knowledge; iv) to inform research policies and institutional strategies; and v) to demonstrate that investment in research is effective and delivers public benefits.

The need for evaluation is fully agreed at the theoretical level, but issues are more problematic when it comes to what methods to apply. The recent development of bibliometric techniques has led various governments to introduce bibliometrics, where applicable, in support or substitution for more traditional peer review. In the United Kingdom the Research Excellence Framework (REF), taking place in 2014, is an informed peer-review exercise, where the assessment outcomes will be a product of expert review informed by citation information and other quantitative indicators. It will substitute the previous Research Assessment Exercise series which were pure peer-review. In Italy, the Quality of Research Assessment (VQR), expected in 2012, substitutes the previous pure peer-review Triennial Evaluation Exercise (VTR). It can be considered a hybrid, as the panels of experts can choose one or both of two methodologies for evaluating any particular output: i) citation analysis; and/or ii) *peer-review* by external experts. The Excellence in Research for Australia initiative (ERA), launched in 2010, is conducted through a pure bibliometric approach for the hard sciences. Single research outputs are evaluated by a citation index referring to world and Australian benchmarks.

The pros and cons of peer-review and bilbiometrics methods have been thoroughly dissected in the literature (Horrobin, 1990; Moxham and Anderson, 1992; MacRoberts and MacRoberts, 1996; Moed, 2002; van Raan, 2005; Pendlebury, 2009; Abramo and D'Angelo, 2011). For evaluation of individual scientific products, the literature fails to decisively indicate whether one method is better than the other but demonstrates that there is certainly a correlation between the results from peer-review evaluation and those from purely bibliometric exercises. This has been demonstrated for the Italian system based on a broad scale study conducted by Abramo et al. (2009), with metrics based on the impact factor of journals, and by Franceschet and Costantini (2011) using citation analysis of publications. Preceding studies concerning other nations have also demonstrated a positive correlation between peer quality esteem and citation indicators (Aksnes and Taxt 2004; Oppenheim and Norris 2003; Rinia et al. 1998; Oppenheim 1997).

The severe limits of peer review emerge when it is applied to comparative evaluation, whether of individuals, research groups or entire institutions. Abramo and D'Angelo (2011) have contrasted the peer-review and bibliometrics approaches in national research assessments and conclude that the bibliometric methodology is by far preferable to peer review in terms of robustness, validity, functionality, time and costs. This is due to the intrinsic limits of all peer-review exercises, in which restrictions on budget and time force the review to focus the evaluation on a limited share of total output from each research organization. One of the consequences is that comparative peer review is limited to the dimension of excellence and is unable to deal with average quality or productivity of the subjects evaluated. A second limitation is that the final rankings are strongly dependent on the share of product evaluated (lack of robustness). A third is that the selection of products to submit to evaluation can be inefficient, due to both technical and social factors (parochialism, the real difficulty of comparing articles from various disciplines, etc.). This can impact negatively on the final rankings and their capacity to represent the true value (or lack of same) for the single organizations evaluated. A fourth consequence is that peer-review evaluations do not offer any assistance to universities in allocating resources to their best individual researchers, since they do not consistently penetrate to precise and comparable levels of information (lack of functionality). Finally, the time and costs of execution involved prevent peer-review evaluations from being sufficiently frequent for effective stimulation of improvement in research systems.

The limitations indicated, particularly those related to the selection and the share of products, lead to legitimate doubts about the accuracy of rankings of organizations as obtained from peer-review national assessment exercises. The aim of this work is to measure the amplitude of shift in rankings of organizations compared to the rankings from bibliometric-type evaluations. Bibliometric simulation is legitimated by the above-noted correlation between peer review and bibliometrics concerning individual research products. The comparison refers to the first Italian research assessment exercise (VTR, 2006), for the scientific production from the period 2001-2003.

The next section of the work describes the dataset used and the methodology for the analysis. Sections 3 and 4 present and comment on the results obtained from the study, conducted at the aggregate level of disciplines. The last section provides a summary of the main results and some further considerations of the authors.

## 2. Methodology

Before showing the comparison between the Italian VTR rankings list[2] and those derived from the bibliometric simulation, we describe the dataset and the specific methodologies applied.

### 2.1 The VTR peer evaluation

In December 2003, the Italian Ministry for Universities and Research (MIUR) launched its first-ever Triennial Research Evaluation (VTR), which for the opening occasion referred to the period 2001-2003. The national Directory Committee for the Evaluation of Research (CIVR) was made responsible for conducting the VTR (2006). The assessment system was designed to evaluate research and development carried out by public research organizations (102 in total), including both universities and research organizations with MIUR funding. However, the remainder of the current work pertains only to universities.

---

[2] http://vtr2006.cineca.it/index_EN.html, last accessed on July 5, 2011.

In Italy each university scientist belongs to one specific disciplinary sector (SDS), 370 in all[3], grouped in 14 University Disciplinary Areas (UDAs). As a first step, the CIVR selected experts for 14 panels, one for each UDA[4]. Universities were then asked to autonomously submit research outputs to the panels[5]: outputs were to be in the proportion of one every four researchers working in the university in the period under observation. Outputs acceptable were limited to articles, books, and book chapters; proceedings of national and international congresses; patents and designs; performances, exhibitions and art works. Thus the VTR was designed as an ex-post evaluation exercise focused on the best outputs produced by Italian research institutions.

In the next step, the panels assessed the research outputs and attributed a final judgment to each product, giving ratings of either "excellent", "good", "acceptable" or "limited". The panels were composed of 183 high level peers appointed by the CIVR, and called on additional support from outside experts. The judgments were made on the basis of various criteria, such as quality, relevance and originality, international scope, and potential to support competition at an international level. To this purpose, the following quality index ($R_{i,u}$) was used for ranking research institution "$i$" in UDA "$u$":

$$R_{i,u} = \frac{1}{T_{i,u}} \cdot (E_{i,u} + 0.8\, G_{i,u} + 0.6\, A_{i,u} + 0.2\, L_{i,u}) \qquad [1]$$

Where:
$E_{i,u}$; $G_{i,u}$; $A_{i,u}$; $L_{i,u}$ = numbers of "excellent, good, acceptable" and "limited" outputs submitted by the $i$th university in UDA $u$
$T_{i,u}$ = total number of outputs submitted by the $i$th university in UDA $u$

A final report ranks universities based on their results under the quality assessment index. The rankings were realized at the level of single UDAs. Within each UDA the universities were subdivided by size into four classes: very large, large, medium, and small. As an example, Table 1 shows the ranking list of Italian "large" universities based on $R_{i,u}$, in the UDA "Mathematics and computer science". Table 1, in addition to the dimensional ranking, gives the excellence ranking within the universe of institutions active in the UDA under examination. Table 2 presents the example of the specific ratings obtained by the University of Rome "Tor Vergata", in the 11 disciplinary UDAs for which it submitted outputs.

The magnitude of the VTR effort can be suggested by a few pertinent facts: the evaluation included 102 research institutions (77 universities and 25 public research organizations) and examined about 18,000 outputs, drawing on 20 peer panels, 183 panelists and 6,661 reviewers, with the work taking almost two years and with direct costs mounting to 3.5 million euros.

---

[3] Complete list accessible at http://cercauniversita.cineca.it/php5/settori/index.php, last accessed on July 5, 2011.
[4] The CIVR also organized six additional panels for "interdisciplinary sectors": Science and technology (ST) for communications and an information society; ST for food quality security, ST for nano-systems and micro-systems; aerospace ST, and ST for the sustainable development and governance.
[5] Each university was also asked to provide the CIVR with sets of input and output data for the institution and its individual UDAs.

| University | Selected outputs | E | G | A | L | Rating | Category rank | Absolute rank | Absolute rank (percentile) |
|---|---|---|---|---|---|---|---|---|---|
| Milan | 28 | 17 | 10 | 1 | 0 | 0.914 | 1 | 4 | 92 |
| Milan Polytechnic | 25 | 16 | 7 | 2 | 0 | 0.912 | 2 | 6 | 90 |
| Pisa | 42 | 22 | 18 | 2 | 0 | 0.895 | 3 | 9 | 85 |
| Rome "La Sapienza" | 61 | 31 | 26 | 4 | 0 | 0.889 | 4 | 13 | 77 |
| Bologna | 35 | 17 | 15 | 3 | 0 | 0.880 | 5 | 16 | 67 |
| Padua | 31 | 11 | 17 | 3 | 0 | 0.852 | 6 | 23 | 58 |
| Florence | 31 | 12 | 15 | 3 | 1 | 0.839 | 7 | 25 | 54 |
| Palermo | 31 | 9 | 14 | 7 | 1 | 0.794 | 8 | 39 | 27 |
| Turin | 30 | 7 | 15 | 7 | 1 | 0.780 | 9 | 41 | 19 |
| Genoa | 30 | 7 | 17 | 4 | 2 | 0.780 | 9 | 41 | 19 |
| Naples "Federico II" | 43 | 7 | 26 | 8 | 2 | 0.767 | 11 | 44 | 17 |

*Table 1: VTR rank list of Italian "large" universities for Mathematics and computer science: E, G, A and L indicate numbers of outputs rated by VTR as excellent, good, acceptable, limited*

| UDA | Selected outputs | E | G | A | L | Rating | Category rank (class) |
|---|---|---|---|---|---|---|---|
| Mathematics and computer science | 23 | 17 | 5 | 1 | 0 | 0.939 | 1 out of 15 (medium) |
| Physics | 19 | 10 | 9 | 0 | 0 | 0.905 | 8 out of 23 (medium) |
| Chemistry | 8 | 3 | 5 | 0 | 0 | 0.875 | 7 out of 26 (small) |
| Biology | 38 | 21 | 13 | 4 | 0 | 0.889 | 5 out of 23 (large) |
| Medicine | 93 | 23 | 51 | 12 | 7 | 0.778 | 10 out of 16 (very large) |
| Civil engineering and architecture | 10 | 2 | 5 | 3 | 0 | 0.780 | 5 out of 15 (medium) |
| Industrial and information engineering | 21 | 5 | 10 | 2 | 4 | 0.714 | 18 out of 18 (medium) |
| Arts and humanities | 23 | 13 | 6 | 3 | 1 | 0.861 | 12 out of 17 (medium) |
| History, philosophy, pedagogy and psychology | 15 | 6 | 7 | 2 | 0 | 0.853 | 2 out of 15 (medium) |
| Law | 28 | 4 | 17 | 5 | 2 | 0.750 | 9 out of 15 (large) |
| Economics and statistics | 18 | 0 | 7 | 4 | 7 | 0.522 | 28 out of 31 (medium) |

*Table 2: VTR ratings for University of Rome "Tor Vergata: E, G, A and L indicate numbers of outputs rated by VTR as excellent, good, acceptable, limited*

## 2.2 The bibliometric dataset

The dataset of scientific products examined in the study is based on the Observatory of Public Research (ORP), derived under license from the Thomson Reuters Web of Science (WoS). ORP provides a census of scientific production dating back to 2001, from all Italian public research organizations (95 universities, 76 research institutions and 192 hospitals and health care research organizations). For this particular study the analysis is limited to universities. Beginning from the raw data of the WoS, and applying a complex algorithm for reconciliation of affiliations and disambiguation of the true identity of the authors, each publication (article, review and conference proceeding) is attributed to the university scientist or scientists that produced it (D'Angelo et al., 2011). Every publication is assigned to a UDA on the basis of the SDS to which the author belongs. A research product co-authored by scientists working in different UDAs is assigned to all these UDAs, and a research product co-authored by scientists working in different universities is assigned to all these universities. The field of observation covers the 2001–2003 triennium and is limited to the hard sciences, meaning eight out of the total 14 UDAs: Mathematics and computer science, Physics, Chemistry, Earth science, Biology, Medicine, Agricultural and veterinary sciences and Industrial and information engineering[6]. In the UDAs thus examined, over the 2001–2003 period, there were an average of 31,924 scientists distributed in 69 universities (Table 3).

| UDA | N. of SDSs | Universities | Research staff |
|---|---|---|---|
| Mathematics and computer sciences | 10 | 59 | 3,006 |
| Physics | 8 | 57 | 2,484 |
| Chemistry | 12 | 58 | 3,057 |
| Earth sciences | 12 | 48 | 1,253 |
| Biology | 19 | 63 | 4,752 |
| Medicine | 50 | 57 | 10,301 |
| Agricultural and veterinary sciences | 30 | 49 | 2,867 |
| Industrial and information engineering | 42 | 60 | 4,204 |
| Total | 183 | 69 | 31,924 |

*Table 3: Universities and research staff in the Italian academic system, by UDA; data 2001-2003*

Overall, in the triennium examined, the research staff of these UDAs achieved 84,289 publications[7]. The products submitted for evaluation in the VTR represented less than 9% of the total portfolio. Table 4 shows the representativity of publications submitted, by UDA.

| UDA | VTR products | VTR ORP-listed publications (a) | Total ORP-listed publications (b) | a/b |
|---|---|---|---|---|
| Mathematics and computer science | 751 | 711 (94.7%) | 6,722 | 10.6% |
| Physics | 626 | 596 (95.2%) | 12,919 | 4.6% |
| Chemistry | 758 | 712 (93.9%) | 8,991 | 7.9% |
| Earth science | 323 | 303 (93.8%) | 3,827 | 7.9% |
| Biology | 1,279 | 1,239 (96.9%) | 8,103 | 15.3% |
| Medicine | 2,644 | 2,574 (97.4%) | 27,577 | 9.3% |
| Agriculture and veterinary science | 617 | 571 (92.5%) | 2,650 | 21.5% |
| Industrial and information engineering | 909 | 807 (88.8%) | 13,500 | 6.0% |
| Total | 7,907 | 7,513 (95.0%) | 84,289 | 8.9% |

*Table 4: Number of publications selected for the VTR by universities in each UDA, and their representativity (period 2001-2003)*

## 3. Evaluation of scientific excellence in universities: VTR versus bibliometric assessment

The VTR provided for evaluation of a number of products from each university proportionate to the number of researchers belonging to each UDA[8]. The underlying objective was clearly to identify and reward the universities on the basis of excellence. However the resulting rankings listings present distortions due to two factors. The first is the inefficiency in selection of the best products on the part of the university, which we have already noted. For this, the rankings lists do not reflect true excellence, but rather that suggested by the products submitted, with the distance from reality depending on the inefficiency of the selection. Abramo et al. (2009) have already quantified the inefficiency related to this problem[9]. The second factor concerns the method of identifying excellence. If an exercise is conceived to measure (and reward) excellence, then the ranking lists that it produces should indicate first place for those universities that produce, under equal availability of resources, a greater quantity of excellent research results (top-down approach). However the VTR, in a pattern that is unavoidable under peer review, evaluated a fixed number of products per university, independently of their real excellence (bottom-up approach). Given the assumption, backed by the literature, that peer review and bibliometrics are of equivalent in their evaluations of individual research products, the bibliometric approach can overcome these limits. Through indicators of impact, it is possible to adopt a top-down approach and at the same time eliminate the inefficiency in selection by the universities.

---
[6] The analysis does not consider Civil engineering and architecture because WoS does not cover the full range of research output for this UDA.
[7] This value includes double counts for publications co-authored by researchers from more than one UDA.
[8] In theory, a university could have submitted products for only one researcher from each UDA.

Using the bibliometric method for the evaluation of excellence, the position of university $i$ in the national ranking list of UDA $u$ derives from the indicator of excellence $I_{i,u}$, defined:

$$I_{i,u} = \frac{\frac{Ne_{i,u}}{Ne_u}}{\frac{RS_{i,u}}{RS_u}} \qquad [2]$$

where

$Ne_{i,u}$ = Number of excellent research products in UDA $u$ authored by scientists of university $i$

$Ne_u$ = Total number of national excellent research products in UDA $u$

$RS_{i,u}$ = Research staff of university $i$ in UDA $u$

$RS_u$ = Total national research staff in UDA $u$

But how can we qualify the excellence of a research product? From a bibliometric point of view, the excellence of a publication is indicated by the citations that it receives from the scientific community of reference. For the aims of the present work we consider an indicator, named the Article impact index (AII), equal to the standardized citations of a publication, i.e. the ratio of citations received by a publication to the median of citations[10] for all Italian publications of the same year and WoS subject category[11]. The distribution of the AII of national publications of a given UDA permits identification of the excellent products on the basis of a given threshold level. We have simulated two scenarios, one in line with international practice and the other in line with the Italian VTR exercise. Consequently the two reference datasets differ in function of the different selection methods for excellent publications: i) consisting of the top 10% of the national publications per AII in each UDA (analogous to international practice); and ii) consisting of the best publications from a UDA in numbers equal to 25% of the total national members of the UDA (analogous to the VTR guidelines).

---

[9] They found that the average percentages of publications selected by universities for the VTR with a bibliometric quality value lower than the median of the national distribution for all of the university's outputs in a UDA varies from a minimum of 3.7% in biology to a maximum of 29.6% for agricultural and veterinary sciences. Other than this last discipline, notable figures also emerge for industrial and information engineering (26.5%) and mathematics and computer science (24.8%) as disciplines in which the selection process results as particularly ineffective. In six out of eight UDAs there were actually universities that submitted all publications with a bibliometric quality indicator lower than the median for the UDA.

[10] Observed as of 30/06/2009, meaning a citation time window between six and eight years, certainly sufficient for the purposes of this work.

[11] A possible alternative would be to standardize to the world average, as frequently observed in the literature. Standardizing citations to the median value rather than to the average, is justified by the fact that distributions of citations are highly skewed (Lundberg, 2007).

For each of these two scenarios, national ranking lists were prepared in each UDA on the basis of indicator $I_{i,u}$. For comparison with the rankings from the VTR, Spearman coefficients of correlation were calculated (Table 5): these result as significant for five UDAs out of eight for scenario A and six out of eight for scenario B. Between the two scenarios, five UDAs are the same: Mathematics and computer sciences, Chemistry, Biology, Medicine, Agricultural and veterinary sciences. For scenario B, the coefficient also results as significant for Industrial and information engineering. Amongst these areas, the coefficients show a non-weak correlation only in Biology. Thus we certainly cannot affirm that the bibliometric evaluation of excellence provides a framework that thoroughly coincides with the results of the evaluation exercise.

| UDA | Scenario A | | | Scenario B | | |
|---|---|---|---|---|---|---|
| | Correlat coeff. | Two-tailed p-value | | Correlat coeff. | Two-tailed p-value | |
| Mathematics and computer sciences | 0.388 | 0.004 | *** | 0.432 | 0.001 | *** |
| Physics | 0.175 | 0.214 | | 0.177 | 0.208 | |
| Chemistry | 0.494 | 0.000 | *** | 0.468 | 0.001 | *** |
| Earth sciences | 0.132 | 0.412 | | 0.118 | 0.463 | |
| Biology | 0.577 | 0.000 | *** | 0.670 | 0.000 | *** |
| Medicine | 0.500 | 0.000 | *** | 0.506 | 0.000 | *** |
| Agricultural and veterinary sciences | 0.407 | 0.029 | ** | 0.384 | 0.040 | ** |
| Industrial and information engineering | 0.325 | 0.031 | | 0.331 | 0.028 | ** |

*Table 5: Spearman correlation between VTR ranking list and bibliometric ranking list*

Given the correlation analysis, it is useful to analyze the shifts between the ranking lists in terms of variation of percentile and quartile. The results for Scenario A are seen in Table 6. The variations are very substantial: in terms of percentiles, the shifts always involve at least 89% of the universities, with average values falling in the range of 20-31 percentile points and medians in the range of 11-25. Maximum shifts are notable, always greater than 67; in four UDAs the maximum shift is actually over 90 percentiles; in Earth sciences and in Medicine there is the extreme circumstance of the university that places first in the VTR rankings coming last in the rankings on the basis of bibliometric indicator for excellence.

| UDA | Univ. | Percentile variations | | | | Quartile variations | | | |
|---|---|---|---|---|---|---|---|---|---|
| | | Var | Max | Aver. | Median | Var | Max | Aver. | Median |
| Mathematics and computer sciences | 53 | 98% | 81 | 26 | 23 | 68% | 3 | 1 | 1 |
| Physics | 52 | 96% | 96 | 30 | 23 | 69% | 3 | 1 | 1 |
| Chemistry | 51 | 94% | 90 | 22 | 18 | 57% | 3 | 1 | 1 |
| Earth sciences | 41 | 98% | 100 | 31 | 25 | 80% | 3 | 1 | 1 |
| Biology | 55 | 96% | 91 | 20 | 15 | 56% | 3 | 1 | 1 |
| Medicine | 46 | 89% | 100 | 22 | 18 | 52% | 3 | 1 | 1 |
| Agricultural and veterinary sciences | 29 | 93% | 89 | 22 | 11 | 45% | 3 | 1 | 0 |
| Industrial and information engineering | 44 | 95% | 67 | 28 | 22 | 61% | 3 | 1 | 1 |

*Table 6: Statistics for shifts between VTR and bibliometric ranking lists (scenario A: excellent publications = top national 10% per UDA)*

The variations by quartile are also very substantial. At least 45% of the universities active in Agricultural and veterinary sciences shift by at least one quartile, and 80% of those active in Biology register such shifts. In the other UDAs, the percentages of universities that make a shift fall between these two extremes. The values for average and median shift are uniform (equal to one quartile), except for Agricultural and veterinary sciences (median nil), as is the value for maximum shift (3 quartiles) for all the eight UDA examined.

The comparison between VTR and bibliometric rankings for scenario B is presented in Table 7: the results are almost a complete match to those from the comparison for scenario A.

Table 8 provides an examination in more detail concerning the distribution of universities for extent of shift, in quartiles. We provide this examination for the example of scenario B. The most striking cases (shifts of 3 quartiles) are seen in three UDAs: Physics (5 universities out of 52), Earth sciences (5 out of 41) and Industrial and information engineering (5 out of 44). In Physics, of the five universities, three drop from first to last quartile and two rise in the opposite direction, with respect to the CIVR evaluation. In Earth sciences and in Industrial and information engineering these numbers are equal to, respectively, two and three.

| UDA | Univ. | Percentile variations | | | | Quartile variations | | | |
|---|---|---|---|---|---|---|---|---|---|
| | | Var | Max | Aver. | Median | Var | Max | Aver. | Median |
| Mathematics and computer sciences | 53 | 100% | 63 | 25 | 25 | 70% | 2 | 1 | 1 |
| Physics | 52 | 96% | 96 | 29 | 23 | 67% | 3 | 1 | 1 |
| Chemistry | 51 | 96% | 92 | 22 | 12 | 51% | 3 | 1 | 1 |
| Earth sciences | 41 | 98% | 100 | 31 | 23 | 68% | 3 | 1 | 1 |
| Biology | 55 | 89% | 65 | 19 | 17 | 62% | 3 | 1 | 1 |
| Medicine | 46 | 91% | 100 | 22 | 17 | 52% | 3 | 1 | 1 |
| Agricultural and veterinary sciences | 29 | 100% | 93 | 23 | 11 | 59% | 3 | 1 | 1 |
| Industrial and information engineering | 44 | 91% | 74 | 28 | 26 | 66% | 3 | 1 | 1 |

*Table 7: Statistics for shifts between VTR and bibliometric ranking lists (scenario B: excellent publications equal to 25% of national FTE research staff per UDA)*

| UDA | Quartile leap | | | | |
|---|---|---|---|---|---|
| | None | 1 | 2 | 3 | Total |
| Mathematics and computer sciences | 16 | 25 | 12 | 0 | 53 |
| Physics | 17 | 14 | 16 | 5 | 52 |
| Chemistry | 25 | 17 | 6 | 3 | 51 |
| Earth sciences | 13 | 16 | 7 | 5 | 41 |
| Biology | 21 | 27 | 6 | 1 | 55 |
| Medicine | 22 | 13 | 10 | 1 | 46 |
| Agricultural and veterinary sciences | 12 | 9 | 5 | 3 | 29 |
| Industrial and information engineering | 15 | 19 | 5 | 5 | 44 |

*Table 8: Numerosity of universities for extent of shift (in quartiles) for each UDA (scenario B)*

We now imagine a division of the rankings into four classes, as in the four research profile classes of universities applied by the last UK Research Assessment Exercise. The Education Funding Council for England (HEFCE) has adopted a performance-based research funding scheme[12] which does not assign any funds to universities that placed in the lowest of these four classes. Universities with an evaluation of their research profile as first class receive (under equal numbers of research staff) three times more funds of universities in the second class, which in turn receive three times as much as those in the third class. If the resource attribution mechanisms for Italian universities were the same as that for the UK HEFCE, in Industrial and information engineering (as an example) three universities would not have received any funds, even though they place first in national rankings according to reliable bibliometric criteria. On the other hand, two universities that place very low in the national bibliometric classification would receive very large quantities of funds on the basis of the VTR, with very evident distortion of the reward system.

---

[12] For detail: http://www.hefce.ac.uk/research/funding/qrfunding/, last accessed on July 5, 2011.

## 4. VTR versus bibliometric productivity assessment

The main limit of the peer-review evaluation method remains that of not being able to compare the research productivity of organizations without excessive costs and times. The consequence of containing costs is the extreme volatility of rankings with variation of the share of product evaluated, as stated above and as measured in a preceding study by Abramo et al. (2010). The authors' opinion is that a system of evaluation and consequent selective funding should embed productivity measurements, which makes evaluation of the total output necessary. In the hard sciences the publications indexed in such bibliometric data bases as WoS or Scopus, represent a meaningful proxy of total output (Moed, 2005), meaning that the bibliometric method permits comparative measurement of productivity. However, if rankings by quality evaluation based on peer review agree with rankings of universities based on productivity, it is evident that no conflict occurs. In this section we test for this occurrence, meaning we verify whether the research institutions evaluated as excellent in terms of quality are also necessarily those that are most efficient in research activities.

As previously, we first do a correlation analysis and then an analysis of the shifts in rankings. We apply a bibliometric indicator of productivity, defining research productivity ($RP_{i,s}$) of University $i$ in SDS $s$ as:

$$RP_{i,s} = \frac{1}{RS_{i,s}} \sum_{j=1}^{N_{i,s}} AII_j \cdot n_{j,i,s} \quad [3]$$

With:

$AII_j$ = article impact index of publication $j$

$n_{j,i,s}$ = fraction of authors of university $i$ and SDS $s$ to total co-authors of publication $j$ (considering, if publication $j$ falls in life science subject categories, the position of each author and the character of the co-authorship, either intra-mural or extra-mural[13]).

$N_{i,s}$ = total number of publications authored by research staff in SDS $s$ of university $i$

$RS_{i,j}$ = Research staff of university $i$ in SDSs $s$

Once the productivity indicator has been measured at the level of SDS we proceed to aggregation at the UDA level, through standardization and weighting of the data for its SDSs. This method limits the distortion typical of aggregate analyses that do not take account of the varying fertility of the SDSs and their varying representation in terms of members in each UDA (Abramo et al., 2008). The research productivity ($RP_{i,u}$) in a general UDA $u$ of a general university $i$ is thus calculated as:

---

[13] If first and last authors belong to the same university, 40% of citations are attributed to each of them; the remaining 20% are divided among all other authors. If the first two and last two authors belong to different universities, 30% of citations are attributed to first and last authors; 15% of citations are attributed to second and last author but one; the remaining 10% are divided among all others.

$$RP_{i,u} = \sum_{s=1}^{n_u} \left( \frac{RP_{i,s}}{RP_s} \cdot \frac{RS_{i,s}}{RS_{i,u}} \right) \qquad [4]$$

With:
$RP_s$ = Average research productivity of national universities in SDS $s$
$RS_{i,u}$ = Research staff of university $i$ in UDA $u$
$n_u$ = number of SDS in UDA $u$

Table 9 presents the Spearman coefficients of correlation for the ranking lists obtained from the VTR and from application of this bibliometric indicator of productivity. The coefficients are statistically significant in only five UDAs out of eight (Mathematics and computer sciences, Chemistry, Biology, Medicine and Industrial and information engineering), but the values indicate a weak correlation between the two rankings. Once again, the results clearly show that the research institutions evaluated through peer review as excellent in terms of quality are not necessarily those that are most efficient in research activities.

| UDA | Coefficient of correlation | Two-tail p-value | |
|---|---|---|---|
| Mathematics and computer sciences | 0.457 | 0.001 | *** |
| Physics | 0.042 | 0.770 | |
| Chemistry | 0.484 | 0.000 | *** |
| Earth sciences | -0.028 | 0.861 | |
| Biology | 0.484 | 0.000 | *** |
| Medicine | 0.375 | 0.010 | *** |
| Agricultural and veterinary sciences | 0.156 | 0.419 | |
| Industrial and information engineering | 0.469 | 0.001 | *** |

*Table 9: Spearman correlation between VTR ranking lists and bibliometric rankings for productivity*

The analysis of the rankings shifts between the two lists (Table 10) shows obvious differences. For quartile rankings, the percentages of universities with shifts vary from a minimum of 53% in Chemistry to a maximum of 77% in Physics[14]. Just as in the analysis for the preceding section, the results are uniform for values of average and median shift (equal to one quartile) and for maximum shift (equal to three quartiles). It should be noted that a shift equal to 3 quartiles means that a university in the top group of rankings by VTR would thus result in the last, or vice versa.

| UDA | Univ. | Percentile variations | | | | Quartile variations | | | |
|---|---|---|---|---|---|---|---|---|---|
| | | Var | Max | Aver. | Median | Var | Max | Aver. | Median |
| Mathematics and computer sciences | 53 | 96% | 81 | 24 | 21 | 68% | 3 | 1 | 1 |
| Physics | 52 | 100% | 88 | 33 | 28 | 75% | 3 | 1 | 1 |
| Chemistry | 51 | 98% | 86 | 23 | 20 | 53% | 3 | 1 | 1 |
| Earth sciences | 41 | 98% | 100 | 33 | 25 | 71% | 3 | 1 | 1 |
| Biology | 55 | 95% | 67 | 24 | 20 | 67% | 3 | 1 | 1 |
| Medicine | 46 | 100% | 96 | 24 | 16 | 57% | 3 | 1 | 1 |
| Agricultural and veterinary sciences | 29 | 100% | 93 | 30 | 25 | 69% | 3 | 1 | 1 |
| Industrial and information engineering | 44 | 98% | 79 | 23 | 16 | 57% | 3 | 1 | 1 |

*Table 10: Statistics for shifts in rankings between VTR ranking lists and bibliometric rankings for productivity*

---

[14] In Table 1 it is also these two areas that are at extreme opposites in terms of differences between bibliometric rating and CIVR rating.

## 5. Conclusions

Both within the scientific community and beyond, there is unanimous agreement that resources for science should be assigned according to rigorous evaluation criteria. However there is a lively debate on which methods should be adopted to carry out such evaluations. The peer-review methodology has long been the most common. This was the approach for the first large-scale evaluation in Italy (VTR), dealing with the 2001-2003 triennium and concluded in 2006. Recently, the agency responsible prepared the guidelines for an updated national evaluation (the VQR), this time on the basis of a seven-year period and a more ample set of products, but still a peer-review type exercise.

For whatever evaluation intended to inform a research funding scheme, the conception must be of a manner to achieve the strategic objectives the policy-maker is proposing. For the Italian VTR, the objective was to identify and reward excellence: in this work we have attempted to verify the achievement of the objective. To do this we compared the rankings lists from the VTR with those obtained from evaluation simulations conducted with analogous bibliometric indicators. The analyses have highlighted notable shifts, the causes of which the authors have amply examined in previous works. The results justify very strong doubts about the reliability of the VTR rankings in representing the real excellence of Italian universities, and raise a consequent worry about the choice to distribute part of the ordinary funding for university function on the basis of these rankings. One detailed analysis by the authors shows that the VTR rankings cannot even be correlated with the average productivity of the universities. Everything seems to suggest a reexamination of the choices made for the first VTR and the proposals for the new VQR. The time seems ripe for adoption of a different approach than peer review, at least for the hard sciences, areas where publication in international journals represents a robust proxy of the research output, and where bibliometric techniques offer advantages that are difficult to dispute when compared to peer review.